\documentclass[numbered]{trbunofficial}

\usepackage{booktabs}
\usepackage{tabularx}
\usepackage{array}
\usepackage{multirow}
\usepackage{enumitem}
\newcommand{\NPkgs}{185}
\newcommand{\NFiles}{1{,}857}
\newcommand{\NIfaces}{786}
\newcommand{\NSubs}{248}
\newcommand{\NPubs}{500}
\newcommand{\NSrv}{38}
\newcommand{\NParams}{1{,}816}
\newcommand{\NConds}{2{,}749}
\newcommand{\NDecision}{1{,}375}
\newcommand{\NValidation}{2{,}274}
\newcommand{\NFlows}{482}
\newcommand{\NCallEdges}{1{,}052}
\newcommand{\NSelected}{740}
\newcommand{\NPone}{214}
\newcommand{\NPtwo}{526}
\newcommand{\NTgtPkgs}{34}
\newcommand{\NTgtFiles}{107}
\newcommand{\NTargets}{740}
\newcommand{\NLLMModels}{2}
\newcommand{\NConditions}{5}

\newcommand{\FuzzBudget}{600}
\newcommand{\FuzzRealized}{60}
\newcommand{\TotalArtifacts}{3{,}700}

\newcommand{\NDisconfirmed}{615}

\newcommand{\NCompiledGpt}{473}

\newcommand{\RepairRounds}{3}
\newcommand{\LinkedGpt}{289}
\newcommand{\LinkedGptNo}{315}

\begin{document}


\title{LLM-Assisted Dynamic Threat Analysis for Attacker-Reachable Software Weaknesses in Autonomous Vehicles}

\TRBauthor*{Md. Wasiul Haque}{Department of Civil, Construction \& Environmental Engineering, The University of Alabama}{mhaque16@crimson.ua.edu}[2009 Smart Communities and Innovation Building (SCIB), 28 Kirkbride Lane,\\
  Tuscaloosa, AL 35487-0288][0009-0007-8417-9261]
\TRBauthor{Sagar Dasgupta, Ph.D.}{Department of Civil, Construction \& Environmental Engineering, The University of Alabama}{sdasgupta@ua.edu}[2009 Smart Communities and Innovation Building (SCIB), 28 Kirkbride Lane,\\
  Tuscaloosa, AL 35487-0288][0000-0001-8491-662X]
\TRBauthor{Mizanur Rahman, Ph.D.}{Department of Civil, Construction \& Environmental Engineering, The University of Alabama}{mizan.rahman@ua.edu}[2007 Smart Communities and Innovation Building (SCIB), 28 Kirkbride Lane,\\
  Tuscaloosa, AL 35487-0288][0000-0003-1128-753X]
\TRBauthor{Md Rayhanur Rahman, Ph.D.}{Department of Computer Science, The University of Alabama}{mrahman87@ua.edu}[3017 Cyber Hall, 248 Kirkbride Lane,\\
  Tuscaloosa, AL 35487-0288][0000-0003-4980-7350]

\AuthorHeaders{Haque, Dasgupta, Rahman, and Rahman}


\maketitle

\section*{Abstract}
\noindent\textbf{Objectives:}~As autonomous vehicles reach public roads, their software becomes safety-critical. A defect reachable from an attacker input can change how a vehicle steers or brakes. Static analysis flags many candidate sites, but confirming one is reachable and exploitable needs executable artifacts whose manual construction is the bottleneck. We ask whether large language models (LLMs) can automate this for the open-source stack Autoware.

\hfill\break%
\noindent\textbf{Methods:}~We perform a compiler-precise static analysis of Autoware
(\NPkgs{} packages), recovering \NDecision{} decision rules,
\NValidation{} validation checks, and \NFlows{} input-to-safety-output flows, from which we derive a weakness taxonomy and a sample of \NTargets{} reachable sites.
For each site, \NLLMModels{} local open-weight LLMs, a no-static-context ablation, and a naive-template baseline generate artifacts. All \TotalArtifacts{}
sets are compiled against the real build under sanitizers, repaired via a compiler in-the-loop stage, and fuzzed when they compile.

\hfill\break%
\noindent\textbf{Findings:}~The principal result is a build-integration failure taxonomy that locates the binding constraint one stage before the fuzzer: 80\% of first-shot compile failures arise from dependency-wiring rather than program logic. The reasoning model compiled 64\% of its harnesses first, against 6\% for the code-specialized model, and repair reached full object-compileability for the reasoning model only by stubbing the real target; consequently, under half of its harnesses reached the fuzzer and all 37 crashes came from a stub. Within budget, no candidate weakness was dynamically confirmed, reflecting this build-integration barrier rather than a benign attack surface.

\hfill\break%
\noindent\textbf{Novelty:}~The first end-to-end feasibility study of LLM-assisted dynamic analysis on a full AV stack, pairing repository-scale static candidate identification with LLM-generated confirmation and pinpointing where the automation fails.

\hfill\break%
\noindent\textbf{Practical Applications:}~Scalable software-safety assurance is a prerequisite for automated driving. Unaided LLM-assisted dynamic analysis is not yet a trustworthy assurance stage; effort is better spent on build integration than on prompt design, while the static analysis already guides safety review.

\newpage

\section{Introduction}\label{sec:intro}

Autonomous vehicles (AVs) are moving from controlled test environments to public roads, making software correctness a matter of transportation safety rather than software engineering alone. A modern automated-driving stack is a large, distributed, real-time system that processes sensor and network data and produces steering, acceleration, and braking commands. As externally supplied values may traverse many interacting components before influencing actuation, a reachable software defect may become a safety hazard for the vehicle, its occupants, and other road users. Assuring software of this scale remains fundamentally difficult~\citep{haque2025security,koopman2016challenges,kalra2016driving}, and functional-safety and automotive-cybersecurity standards require developers to identify hazardous behavior and demonstrate that it cannot be triggered under relevant operating and threat conditions~\citep{iso26262,isosae21434}.

A central assurance challenge is the gap between identifying a potential weakness and confirming it at run time. Static analysis of the source code can examine an entire repository and locate sites where externally influenced values reach safety-relevant decisions, validation is weak or absent, or data propagate toward consequential outputs \cite{haque2025security}. Such sites are candidates, not confirmed findings. Static evidence alone does not establish runtime reachability, path feasibility, or a security- or safety-relevant effect \cite{aggarwal2006integrating}.

Closing this gap requires an executable artifact that constructs the required state, delivers adversarial inputs to the flagged code, and observes the result \cite{aggarwal2006integrating}. The main obstacle is often not the fuzzer or sanitizer, both of which are mature~\citep{fioraldi2020aflpp,serebryany2012asan,libfuzzer,manes2019fuzzing}, but the construction of the harness itself. A valid harness must satisfy target types, initialization requirements, message formats, middleware dependencies, package boundaries, and build configuration. In AV stacks, generated middleware artifacts like Robot Operating System (ROS)~2 \cite{macenski2022ros2} interfaces, lifecycle assumptions, and cross-package dependencies make this process particularly difficult. Recent work shows that large language models (LLMs) can generate fuzz drivers, test programs, and structured inputs with varying success~\citep{deng2023titanfuzz,xia2024fuzz4all,zhang2024fuzzdriver,ossfuzzgen,hou2024llmse}. This motivates the central question of the study: can an LLM automate artifact construction for a production-scale automated-driving stack and enable dynamic confirmation at a scale impractical for manual effort?

We investigate this question through an end-to-end feasibility study of Autoware \cite{autowareDocs}. We first identify sites where externally influenced data interact with safety-relevant decisions, validation logic, or output paths, and use this evidence to characterize the attack surface and derive a weakness taxonomy. From this population, we select a stratified sample and provide the associated code and static context to two local, open-weight LLMs. Each model generates an executable artifact for the selected target. The artifacts are compiled and linked against the native Autoware build with sanitizers enabled. Failed artifacts enter a compiler-in-the-loop repair process, and successful ones are executed under a fixed fuzzing budget. This design evaluates the full path from static candidate identification to dynamic confirmation while exposing the intermediate failures that prevent execution.
The study addresses the following research questions:

\emph{\textbf{RQ1:}} What classes of potentially exploitable weaknesses are reachable from external inputs in a production-scale automated-driving stack, and can LLM-generated dynamic artifacts confirm those weaknesses within a fixed analysis budget?

\emph{\textbf{RQ2:}} To what extent can LLM-assisted dynamic analysis refine, correct, or extend the conclusions produced by static analysis alone, and what factors prevent it from doing so?

\emph{\textbf{RQ3:}} How does model choice affect artifact validity, target-code reachability, and the severity of confirmed findings, and, when no findings are confirmed, how does it affect the intermediate outcomes, including compileability, failure mode, and generation cost that determine whether dynamic analysis can proceed?

In our experiment, the results provide a negative but informative assessment of current feasibility. No candidate was dynamically confirmed within the allocated fuzzing budget, primarily because most generated artifacts failed before meaningful execution rather than because Autoware's attack surface was shown to be benign. Unaided LLM outputs rarely compiled and linked against the native build, while compiler-guided repair raised object compileability to 100\% for the stronger model largely by introducing stubs or replacement implementations that bypassed the intended target. The study therefore identifies faithful build integration and verified target execution, rather than input generation alone, as the principal barrier to LLM-assisted dynamic confirmation. Based on this investigation, the paper makes the following contributions:
\begin{itemize}[leftmargin=*]
\item We provide a repository-scale, compiler-precise characterization of an open-source AV stack, Autoware's safety-relevant attack surface, comprising \NDecision{} decision rules, \NValidation{} validation checks, and \NFlows{} input-to-safety-output paths across \NPkgs{} packages. From this evidence, we derive a weakness taxonomy grounded in observed Autoware code patterns rather than in a generic vulnerability catalog.
\item We develop an end-to-end and reproducible pipeline that carries each static candidate through LLM-based artifact generation, integration with the native Autoware build, sanitizer-enabled compilation, compiler-guided repair, fixed-budget fuzzing, and post-execution triage.
\item We conduct a controlled comparison of \NLLMModels{} LLMs, together with a no-static-context ablation and a naive baseline, to determine how model capability and contextual information affect artifact construction and target-code reachability.
\item We derive a build-integration failure taxonomy that identifies the stages at which LLM-assisted dynamic analysis breaks down on a codebase of this scale and distinguishes genuine target execution from artifacts that achieve nominal compilation by replacing, bypassing, or stubbing the intended implementation.
\end{itemize}

The remainder of the paper is organized as follows. Section~\ref{sec:related} reviews related work on software assurance, fuzzing, and LLM-assisted testing. Section~\ref{sec:stacks} surveys automated-driving software stacks and motivates the selection of Autoware, while Section~\ref{sec:threat} defines the threat model. Section~\ref{sec:static} presents the static analysis and weakness taxonomy. Section~\ref{sec:method} describes the analysis pipeline, and Section~\ref{sec:setup} presents the experimental setup, reports the results, followed by their interpretation in Section~\ref{sec:discussion}. Section~\ref{sec:conclusion} discusses limitations, threats to validity, and concludes the paper.

\section{Related Work}\label{sec:related}
Confirming security-relevant weaknesses in automated-driving software requires several complementary capabilities: understanding how untrusted inputs propagate through the stack, identifying suspicious code paths at repository scale, exercising those paths dynamically, and constructing the artifacts needed to make such execution possible. This section reviews the corresponding foundations in automated-driving software assurance, static analysis and fuzzing, and LLM-assisted software testing, and positions the present study at their intersection.

\subsection{Automated-Driving Software and Safety Context}
Modern AV stacks are distributed, real-time systems spanning perception, localization, prediction, planning, control, and map processing. Open platforms commonly use middlewares like ROS~2, where nodes exchange serialized messages through a Data Distribution Service (DDS)-based publish--subscribe model~\citep{macenski2022ros2}. Deserialization and subscription callbacks mark the point where external bytes become typed C++ objects, making them natural entry points for both attacker-controlled data and fuzzing inputs. System behavior also depends on launch files, parameters, topic remappings, and cross-package dependencies, allowing inputs to traverse several domains before affecting control. SAE~J3016 defines automation levels~\citep{saej3016}, ISO~26262 addresses functional safety~\citep{iso26262}, and ISO/SAE~21434 addresses vehicle cybersecurity~\citep{isosae21434}. Because the behavioral space cannot be covered by road testing alone~\citep{haque2025security, koopman2016challenges,kalra2016driving}, scalable software-level assurance is required.

\subsection{Static Analysis, Fuzzing, and Harness Construction}
The static stage builds on sound program analysis~\citep{cousot1977abstract} and 
Low Level Virtual Machine (LLVM)/Clang~\citep{lattner2004llvm,clangtooling}. Clang analyzes each translation unit under its actual build configuration, including generated interfaces, include paths, and preprocessor definitions. Tools like CodeQL provide a complementary repository-scale approach to control- and data-flow analysis~\citep{codeqlOverview}. Static findings remain candidates because reported paths may be infeasible, unreachable, or harmless at run time. Coverage-guided fuzzing and sanitizers provide the mechanisms needed to test them~\citep{manes2019fuzzing,fioraldi2020aflpp,libfuzzer,serebryany2012asan}, and continuous-fuzzing systems demonstrate their scalability when suitable targets exist~\citep{serebryany2017ossfuzz}. These tools require an executable harness that compiles, links, initializes the target, and carries fuzzer-controlled data into the intended implementation. In Autoware, this may involve generated ROS~2 types, node state, package dependencies, and middleware initialization. Classical fuzz-driver synthesis can infer Application Programming Interface (API) usage from examples~\citep{ispoglou2020fuzzgen}, but is less effective for targets embedded in distributed node architectures and large build graphs.

\subsection{LLMs for Software Testing and Security}

LLMs are increasingly used for code generation, debugging, testing, vulnerability analysis, and repair~\citep{hou2024llmse}. In fuzzing, they have generated programs and inputs for deep-learning frameworks and general software~\citep{deng2023titanfuzz,xia2024fuzz4all}, guided protocol fuzzing~\citep{meng2024chatafl}, drafted continuous-fuzzing harnesses~\citep{ossfuzzgen}, and generated library fuzz drivers~\citep{zhang2024fuzzdriver}. LLM agents have also constructed exploits from known vulnerability descriptions~\citep{fang2024llmexploit}. Many approaches use iterative repair: the artifact is compiled, diagnostics are returned, and the model revises its output. Our compiler-in-the-loop process follows this reasoning-and-acting pattern~\citep{yao2023react}. However, compilation alone does not ensure semantic fidelity. A model may remove dependencies, replace implementations, or introduce stubs that bypass the real target. These artifacts cannot confirm the original candidate.

\subsection{Positioning of This Study}
Most LLM-based fuzz-driver studies evaluate isolated programs, libraries, or APIs. Our study integrates repository-scale static analysis, candidate-guided LLM generation, native-build compilation and repair, fuzzing, and target-reachability verification. This design distinguishes failures in candidate selection, generation, build integration, and execution. To our knowledge, it is the first end-to-end feasibility study of LLM-assisted dynamic analysis on a complete autonomous driving stack that treats build integration and real target execution as primary outcomes.

\section{Autonomous Vehicle Software Stacks and the Study Subject}\label{sec:stacks}
\begin{figure}[htbp]
\centering
\includegraphics[width=0.86\linewidth]{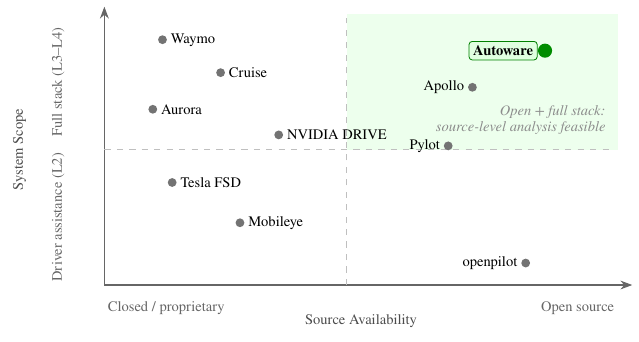}
\caption{Automated-driving software stacks positioned by source availability and system scope. Autoware, highlighted in the figure, is selected as
the subject of this study.}
\label{fig:stacks}
\end{figure}

Autonomous driving software ranges from proprietary commercial systems to fully open-source platforms~\citep{yurtsever2020survey}. Because our method requires source code, compiler metadata, package dependencies, and the native build environment, only open platforms are suitable. Table~\ref{tab:stacks} and Figure~\ref{fig:stacks} summarize this landscape. Commercial systems such as Waymo, Cruise, Aurora, Tesla FSD, Mobileye, and NVIDIA DRIVE are proprietary. Waymo has published elements of its Level~4 safety-case methodology~\citep{webb2020waymo}, while Mobileye introduced the Responsibility-Sensitive Safety model~\citep{shalevshwartz2017rss}; however, none exposes the source and build system required for our analysis. Among open platforms, Autoware and Apollo provide the most complete deployment-oriented stacks. Autoware is a ROS~2-based Level~3--4 platform maintained by the Autoware Foundation and released under Apache~2.0~\citep{kato2018autoware,autowareDocs}. Apollo is a similarly comprehensive Apache-licensed stack~\citep{apollo,jung2025autowareapollo}. openpilot is limited to Level~2 driver assistance~\citep{openpilot}, while Pylot is primarily a modular research platform~\citep{gog2021pylot}.

\begin{table}[!ht]
  \caption{Representative Automated-Driving Software Stacks}\label{tab:stacks}
  \begin{center}\footnotesize
  \begin{tabularx}{\linewidth}{lllX}
    \toprule
    Stack & Developer & Access & Scope \\
    \midrule
    Waymo Driver & Waymo (Alphabet) & Proprietary & Level~4 Robotaxi \\
    Cruise & General Motors & Proprietary & Level~4 Robotaxi \\
    Aurora Driver & Aurora & Proprietary & Level~4, Autonomous trucking \\
    Tesla FSD & Tesla & Proprietary & Vision-centric Level~2, Fleet scale \\
    Mobileye & Mobileye (Intel) & Proprietary & Camera-based ADAS to Level~4; RSS model \\
    NVIDIA DRIVE & NVIDIA & Proprietary & Compute and software SDK for AD \\
    \midrule
    Autoware & Autoware Foundation & Open (Apache 2.0) & ROS~2 full Level~3-4 stack \\
    Apollo & Baidu & Open (Apache 2.0) & Full stack; Widely used \\
    openpilot & comma.ai & Open (MIT) & Level~2 driver assistance on consumer cars \\
    Pylot & Research & Open (Apache 2.0) & Modular research AD platform \\
    \bottomrule
  \end{tabularx}
  \end{center}
\end{table}

Source and build-system availability were essential because the pipeline analyzes compiler representations and compiles generated artifacts against the native implementation, excluding proprietary platforms. System scope further narrowed the candidates to complete automated-driving stacks, favoring Autoware and Apollo over narrower platforms such as openpilot and Pylot. Autoware was ultimately chosen for its compatibility with the proposed workflow. Its ROS~2 architecture exposes message interfaces, package boundaries, launch configurations, dependencies, and a reproducible \texttt{colcon}-based build with package-level compile databases.

\section{Threat Model}\label{sec:threat}
We consider a software-level adversary capable of influencing one or more inputs to the Autoware node graph. Attacks requiring physical access to the vehicle, sensors, or environment are out of scope. The adversary may act through four channels aligned with the input sources traced in our static analysis: (i) the adversary may inject a malicious or spoofed sensor stream into a perception or localization topic; (ii) a rogue DDS participant may join the ROS~2 graph and publish to or subscribe from safety-relevant topics, without an enforced security policy, DDS discovery may admit previously unknown participants~\citep{macenski2022ros2,dieber2017security}; (iii) a compromised upstream node may emit well-typed but semantically adversarial messages, including stale, inconsistent, out-of-range, non-finite, or boundary values; and (iv) the adversary may manipulate map, route, or localization inputs consumed at startup or during execution.

\begin{figure}[htbp]
  \centering
  \includegraphics[width=\linewidth]{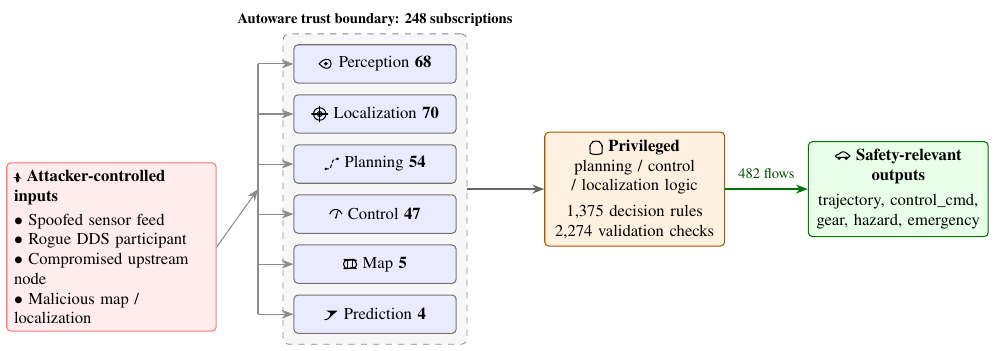}
  \caption{Trust boundary over the Autoware node graph, derived from the static
  interface inventory. Attacker-controlled inputs reach each domain's subscription
  surface, which feeds privileged decision logic and safety-relevant outputs.}
  \label{fig:trust}
\end{figure}

The adversary aims to cause denial of service in a safety-critical node, corrupt perception or localization state, induce unsafe planning or control decisions, or silently degrade motion-related outputs without producing an immediate crash. Figure~\ref{fig:trust}, derived from the static ROS~2 interface inventory, illustrates how these inputs cross into perception, localization, planning, and control logic before reaching safety-relevant outputs. We exclude physical sensor spoofing and adversarial perturbations against learned perception models~\citep{cao2019adversarial,sato2021dirty}, supply-chain compromise of code, dependencies, models, or build artifacts, and attacks on V2X or roadside infrastructure. All experiments are conducted in software-in-the-loop.

\section{Safety-Relevant Attack Surface of Autoware}\label{sec:static}
Before dynamic confirmation, we identify where safety-relevant weaknesses may plausibly occur. This section presents the repository-scale static analysis used to characterize Autoware's attack surface, select candidates for dynamic testing, and derive the weakness taxonomy. All reported counts and figures are generated by the analysis pipeline and released with the accompanying artifacts.

\subsection{Analysis Method and Scale}
Whole-repository extraction identifies packages, domains, deployed sources, dependencies, and ROS interfaces while excluding non-deployment code. Clang then processes each translation unit under its exported build configuration to recover \texttt{if} and \texttt{switch} conditions and a simplified call graph. Launch and YAML files provide node composition, remappings, and runtime configuration, with CodeQL used as a complementary cross-check~\citep{codeqlOverview}. Finally, classifiers label safety-relevant decisions and validation checks, and a heuristic package-level analysis links input subscriptions to consequential outputs.

\begin{figure}[ht]
  \centering
  \includegraphics[width=\linewidth]{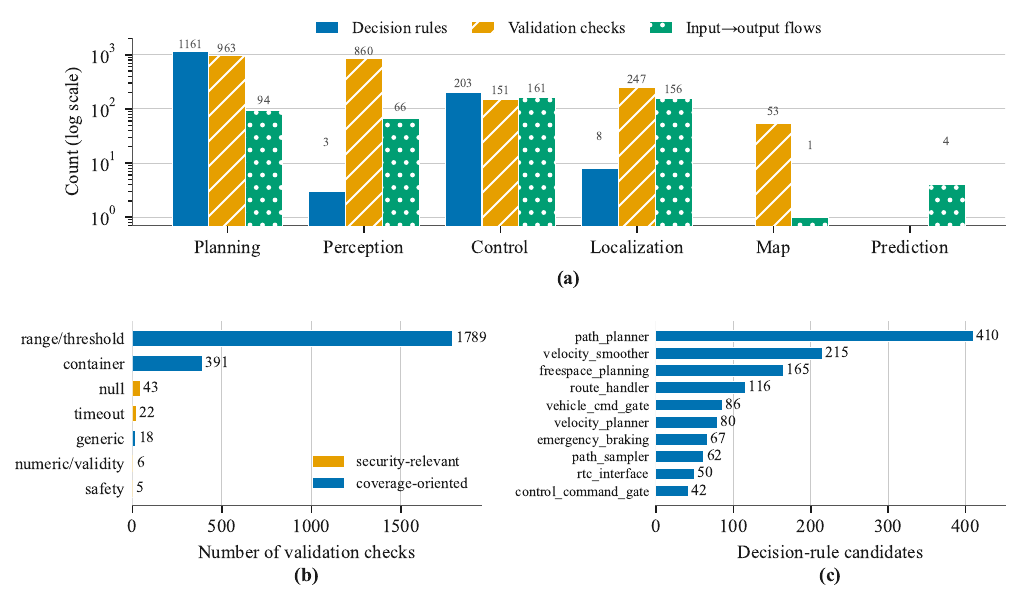}
  \caption{Autoware's safety-relevant static surface. (a)~Decision rules, validation checks,
  and heuristic input-to-output flows by domain (log scale) (b)~Validation checks by type (c)~The ten most decision-dense packages.}
  \label{fig:surface}
\end{figure}

The pipeline analyzes \NPkgs{} packages and \NFiles{} source files as listed in Table~\ref{tab:staticscale}. It recovers \NIfaces{} ROS interfaces, comprising \NSubs{} subscriptions, \NPubs{} publishers, and \NSrv{} services, together with \NParams{} parameter surfaces and \NConds{} compiler-precise condition sites. Figure~\ref{fig:surface}(a) summarizes the recovered surface.

\begin{table}[ht]
  \caption{Repository-Scale Static-Analysis Inventory for Autoware}\label{tab:staticscale}
  \begin{center}\footnotesize
  \begin{tabular}{lr}
    \toprule
    Quantity & Count \\
    \midrule
    Relevant packages & \NPkgs{} \\
    Source files in scope & \NFiles{} \\
    ROS interfaces (subs / pubs / services) & \NIfaces{} (\NSubs{}/\NPubs{}/\NSrv{}) \\
    Parameter surfaces & \NParams{} \\
    Compiler-precise condition sites & \NConds{} \\
    Decision-rule candidates & \NDecision{} \\
    Validation checks & \NValidation{} \\
    Heuristic input-to-output paths & \NFlows{} \\
    Call-graph edges & \NCallEdges{} \\
    \bottomrule
  \end{tabular}
  \end{center}
\end{table}

\subsection{Structure of the Attack Surface}
Safety-relevant decision logic is abundant and distributed: the pipeline recovers \NDecision{} candidates, concentrated in planning (1{,}161) and control (203), across route management, velocity regulation, command gating, emergency braking, and shared planning infrastructure. The heuristic analysis recovers \NFlows{} package-level paths, concentrated in the vehicle command gate, external command selector, and control command gate, with representative examples in Table~\ref{tab:flows}. Because this analysis is not fully interprocedural or taint-precise, it supports prioritization rather than proof of attacker-to-actuation reachability. Validation is also uneven: among \NValidation{} recovered checks, 1{,}789 are range or threshold guards and 391 are container checks, compared with 43 null checks, 22 timeout or staleness checks, and 6 numeric-validity checks. These are summarized in Figures \ref{fig:surface}(b) and \ref{fig:surface}(c).


\begin{table}[htbp]
  \caption{Representative Attacker-Influenced Flows from Input to Safety-Relevant Output}\label{tab:flows}
  \begin{center}\footnotesize
  \begin{tabularx}{\linewidth}{lX}
    \toprule
    Domain & Input topic $\rightarrow$ package $\rightarrow$ safety output \\
    \midrule
    Control & \texttt{input/external\_emergency\_stop\_heartbeat} $\rightarrow$ \texttt{vehicle\_cmd\_gate} $\rightarrow$ \texttt{output/vehicle\_cmd\_emergency} \\
    Planning & \texttt{\textasciitilde/input/odometry} $\rightarrow$ \texttt{mission\_planner} $\rightarrow$ \texttt{output/goal\_pose} \\
    Perception & \texttt{\textasciitilde/input/image} $\rightarrow$ \texttt{bevfusion} $\rightarrow$ \texttt{\textasciitilde/output/objects} \\
    Localization & \texttt{\textasciitilde/input/pose\_with\_covariance} $\rightarrow$ \texttt{yabloc\_particle\_filter} $\rightarrow$ \texttt{\textasciitilde/output/weighted\_particles} \\
    \bottomrule
  \end{tabularx}
  \end{center}
\end{table}


Table~\ref{tab:taxonomy} organizes the recovered evidence into four operational weakness classes used for candidate selection, dynamic labeling, and severity analysis under the adapted CVSS/RVSS framework~\citep{cvss31,vilches2018rvss}. The taxonomy does not classify every site as vulnerable; it identifies the mechanism through which externally influenced data may affect safety-relevant behavior.

\begin{table}[htbp]
  \caption{Weakness Taxonomy Derived from the Static Analysis}\label{tab:taxonomy}
  \begin{center}\footnotesize
  \begin{tabularx}{\linewidth}{p{0.24\linewidth}Xr}
    \toprule
    Category & Description & Static basis \\
    \midrule
    Decision-policy guard & Conditional controlling a safety-relevant branch (stop,
    yield, emergency, gate-mode, geometry). & \NDecision{} rules \\
    Validation weakness & Missing or weak range, container, null/optional, timeout, or
    numeric-validity guard on an externally influenced input. & \NValidation{} checks \\
    Input dependency & Site fed by an attacker-influenced subscription or parameter close
    to an actuation output. & \NFlows{} flows \\
    State dispatch & \texttt{switch} or mode selection over an externally influenced state
    variable. & subset of conditions \\
    \bottomrule
  \end{tabularx}
  \end{center}
\end{table}

\section{LLM-Assisted Dynamic Analysis Approach}\label{sec:method}
\begin{figure}[tp]
  \centering
  \includegraphics[height=0.9\textheight]{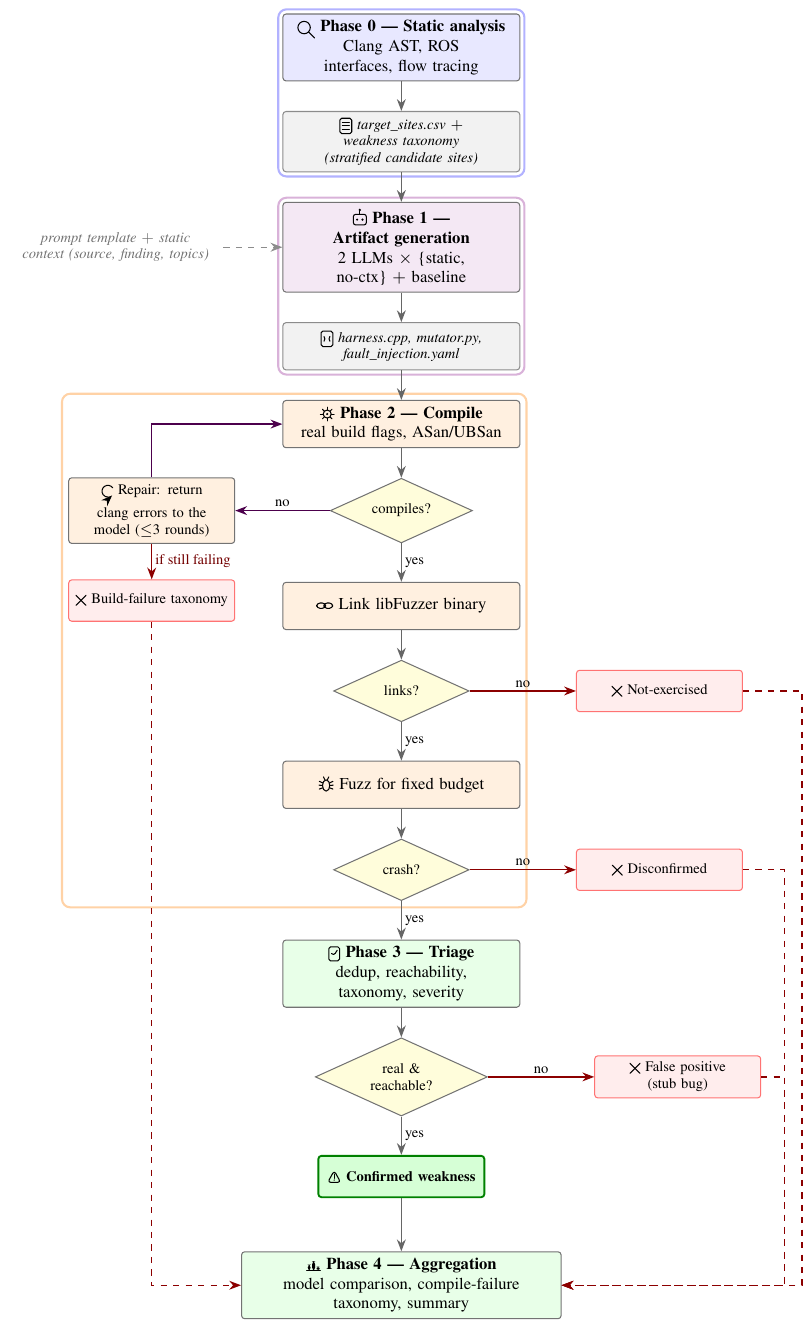}
    \caption{End-to-end LLM-assisted dynamic-analysis pipeline. Solid arrows show progression through the five phases, the purple loop denotes compiler-guided repair, and decision nodes classify each harness as a build failure, not exercised, disconfirmed, false positive, or confirmed weakness before Phase~4 aggregation.}
  \label{fig:pipeline}
\end{figure}

The pipeline spans five phases, as illustrated in Figure~\ref{fig:pipeline}: (i) static candidate identification
(Phase~0); (ii) artifact generation (Phase~1); (iii) build
integration and dynamic execution (Phase~2); (iv) triage of each execution outcome (Phase~3); and (v)
aggregation across conditions (Phase~4). This section describes static candidate identification in Phase~0 and the dynamic workflow spanning Phases~1--4. To support a controlled comparison, the prompt template, target context, decoding parameters, build environment, and execution budget are held constant across conditions. The model is the only varying factor, except in the ablation condition, where the static-analysis context is withheld.

\subsection{Target Selection}
The dynamic study evaluates the full set of high-priority candidates identified by the static analysis rather than a manually selected sample. Section~\ref{sec:static} recovers \NConds{} compiler-precise condition sites across \NPkgs{} packages and assigns each a safety-relevance score from 0 to 11. The score rewards actuation-related terms (+3), validation or bounds checks (+2), external ROS input exposure (+2), presence on an input-to-safety-output flow (+3), and membership in a safety-critical domain (+1). Sites scoring ($\ge 8$) are classified as P1, those scoring 5--7 as P2, and the remainder as P3. All P1 and P2 sites are retained, yielding \NSelected{} targets: \NPone{} in P1 and \NPtwo{} in P2, spanning \NTgtPkgs{} packages and \NTgtFiles{} source files. Table~\ref{tab:sample} presents the composition by domain and tier. The selection is deterministic, reproducible, and exhaustive within these tiers.

\begin{table}[!ht]
  \caption{Composition of the \NSelected{}-Target Census by Domain and Priority Tier
  (Every P1 and P2 Site; P3 Excluded)}\label{tab:sample}
  \begin{center}\footnotesize
  \begin{tabular}{lrrr}
    \toprule
    Domain & P1 & P2 & Total \\
    \midrule
    Planning     & 129 & 232 & 361 \\
    Control      &  59 &  79 & 138 \\
    Perception   &   6 & 118 & 124 \\
    Localization &  20 &  95 & 115 \\
    Map          &   0 &   2 &   2 \\
    \midrule
    Total        & \NPone{} & \NPtwo{} & \NSelected{} \\
    \bottomrule
  \end{tabular}
  \end{center}
\end{table}

\subsection{Artifact Generation}
\begin{figure}[htbp]
  \centering
  \includegraphics[width=\linewidth]{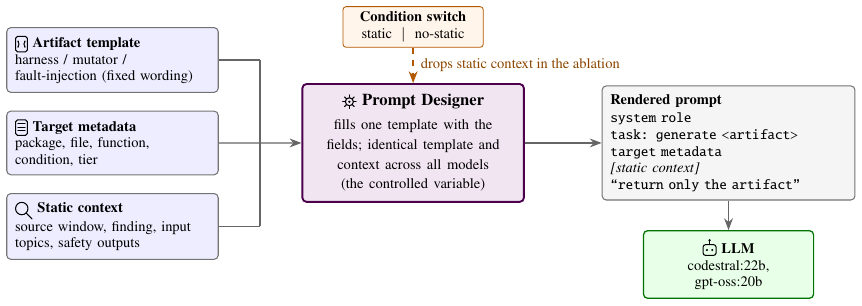}
  \caption{The prompt designer for Phase 1}
  \label{fig:prompt}
\end{figure}

For each target, a fixed prompt template is populated with the target metadata and static-analysis context. The context includes the relevant source window, the candidate description, and the recovered input topics and safety-relevant outputs, as illustrated in Figure~\ref{fig:prompt}. The template, context fields, and decoding parameters are identical across models. The no-static-context ablation receives only target metadata, while a deterministic baseline produces metadata-based scaffolds as a naive lower bound. All prompts and raw responses are retained for reproducibility.

Each target produces four artifacts: (i) a function-level libFuzzer harness, (ii) a ROS~2 message mutator, (iii) a software-in-the-loop fault-injection specification, and (iv) an ASan/UBSan build configuration. The mutator generates malformed, boundary, and timing-variant messages, while the fault-injection specification defines bounded message drops, delays, stale replays, and NaN/Inf perturbations.

\subsection{Build Integration and Repair}
Each harness is compiled against the pinned Autoware revision using the target translation unit's include paths, preprocessor definitions, and compiler options from \texttt{compile\_commands.json}. Compilation uses \texttt{clang++} with AddressSanitizer and UndefinedBehaviorSanitizer. We define compileability as successful object compilation against the real Autoware headers and APIs, a less stringent criterion than linking or execution. All attempts are logged and classified using a fixed error taxonomy. Failed harnesses enter a compiler-in-the-loop repair process for up to \RepairRounds{} rounds. The model receives the failing source and Clang diagnostics, while the original output is retained to distinguish first-shot from post-repair compileability. In the no-static-context condition, repair uses only the generated artifact and compiler feedback; the withheld static context is not reintroduced.

\subsection{Triage and Aggregation}
Compiled harnesses are linked into libFuzzer executables and run under a fixed budget. A harness is considered valid only if it invokes the intended Autoware implementation and carries fuzzer-controlled input to the target. Crashes are deduplicated and manually assessed for reachability, reproducibility, and security or safety impact. A candidate is classified as confirmed when the real target is exercised and a reproducible weakness is observed, and as a false positive when the crash originates outside the target. A target exercised for the full budget without a relevant failure is disconfirmed within that budget, whereas a harness that does not reach the intended code or fuzzing stage is not exercised. This distinction prevents build and integration failures from being treated as evidence against the static candidate.

\section{Evaluation And Results}\label{sec:setup}
This section reports the end-to-end outcomes for the \NTargets{} targets selected in Section~\ref{sec:static}. Across the \NConditions{} conditions defined in Section~\ref{sec:setup}, the pipeline generated \TotalArtifacts{} artifact sets, compiled and linked them against the sanitized Autoware build, repaired failures through the compiler-in-the-loop process, and fuzzed the surviving harnesses. Because the results are dominated by failures before execution, we organize them by pipeline stage rather than by research question.

\subsection{Model Configuration and Parameters}

The experimental setup is designed to compare model behavior under a common execution environment. We evaluate two local, open-weight LLMs served through an Ollama OpenAI-compatible endpoint: one code-specialized model and one general reasoning model. Using models with different capabilities helps separate model-specific effects from those introduced by the prompt and pipeline design. All experiments target the same Autoware workspace and use a configured fuzzing budget of \FuzzBudget{} seconds per target. For harnesses that successfully linked, the realized execution time was \FuzzRealized{} seconds per target. Table~\ref{tab:setup} summarizes the complete configuration.

\begin{table}[!ht]
  \caption{Experimental Configuration}\label{tab:setup}
  \begin{center}\footnotesize
  \begin{tabularx}{\linewidth}{lX}
    \toprule
    Field & Value \\
    \midrule
    Target subject & Autoware (cloned on 10 February 2026) \\
    Models (LLM) & \texttt{codestral:22b} (code-specialized); \texttt{gpt-oss:20b} (open-weight reasoning) \\
    Decoding & temperature 0.1, max\_tokens 4096 \\
    Conditions & 2 models $\times$ \{static, no-static\} + naive baseline \\
    Build repair & compiler-in-the-loop, up to \RepairRounds{} rounds \\
    Hardware & Intel i9-14900F (32 threads); NVIDIA RTX 4090 (24\,GB) \\
    Sample & \NTargets{} sites, stratified by domain $\times$ priority, seed 202707 \\
    Sanitizers & AddressSanitizer + UndefinedBehaviorSanitizer \\
    \bottomrule
  \end{tabularx}
  \end{center}
\end{table}

A harness provides evidence about Autoware only if it compiles against the target package, links to the real implementation rather than substituted stubs, and exercises the flagged branch at run time. Only then can a crash, or its absence, be attributed to the target code. We therefore present the results as attrition through these gates: build integration and target reachability, which address RQ1 and RQ2; the effects of model choice and compiler-guided repair, which address RQ3; and the evidence provided by the few harnesses that reach execution.

\begin{table}[!ht]
  \caption{Compile-integration outcomes and first-shot compile-failure classes by condition
  (\NTargets{} harnesses per LLM condition). Columns $n$ and \% give the total and share across
  the four LLM conditions.}\label{tab:comp}
  \begin{center}\footnotesize
  \setlength{\tabcolsep}{4pt}
  \renewcommand{\arraystretch}{1.15}
  \begin{tabularx}{\linewidth}{@{}Xrrrrrr r@{}}
    \toprule
     & \multicolumn{4}{c}{LLM conditions} & \multicolumn{2}{c}{LLM (all four)} & \\
    \cmidrule(lr){2-5}\cmidrule(lr){6-7}
     & codestral & \shortstack{codestral\\(no ctx)} & gpt-oss & \shortstack{gpt-oss\\(no ctx)}
       & $n$ & \% & baseline \\
    \midrule
    \multicolumn{8}{@{}l}{\itshape Compile-integration outcome}\\
    \quad Compiled, first-shot   & 46 & 3  & 473 & 179 & 701 & 24 & 740 \\
    \quad Compiled, after repair & 95 & 12 & 740 & 728 & 1{,}575 & 53 & 740 \\
    \quad Linked \& fuzzed       & 38 & 10 & 289 & 315 & 652 & 22 & --- \\
    \addlinespace
    \multicolumn{8}{@{}l}{\itshape First-shot compile-failure class}\\
    \quad Missing include        & 441 & 536 & 62 & 397 & 1{,}436 & 64 & 0 \\
    \quad Source file not found  & 134 & 120 & 21 & 106 & 381 & 17 & 0 \\
    \quad API mismatch           & 64  & 53  & 65 & 43  & 225 & 10 & 0 \\
    \quad Signature mismatch     & 37  & 22  & 44 & 7   & 110 & 5  & 0 \\
    \quad Syntax error           & 8   & 4   & 62 & 7   & 81  & 4  & 0 \\
    \quad Other                  & 10  & 2   & 13 & 1   & 26  & 1  & 0 \\
    \cmidrule(lr){1-8}
    \quad Total first-shot failures & 694 & 737 & 267 & 561 & 2{,}259 & 100 & 0 \\
    \bottomrule
  \end{tabularx}
  \end{center}
\end{table}

\subsection{Build Integration Analysis (RQ1, RQ2)}
No weakness was confirmed across the \NConditions{} conditions and \NTargets{} targets. Given the thin validation and plausible actuation-facing paths identified in Section~\ref{sec:static}, this result reflects failure before meaningful fuzzing rather than evidence of a benign attack surface. Of the 2,960 LLM-generated harnesses, 2,259 failed to compile initially. As shown in Table~\ref{tab:comp}, 1,436 omitted required ROS or Autoware headers and 381 included target \texttt{.cpp} files through invalid paths, together accounting for 1,817 of 2,259 failures. The remainder comprised API mismatches (225), signature mismatches (110), syntax errors (81), and 26 other errors. Build integration therefore prevented most artifacts from confirming or refuting their static candidates. Table~\ref{tab:comp} also shows distinct model behavior: the code-specialized model failed mainly on dependency wiring, whereas the reasoning model more often reached real APIs but failed on call or signature compatibility.

\subsection{Model Choice, Repair, and Stub Convergence Analysis (RQ3)}

\begin{figure}[htbp]
  \centering
  \includegraphics[width=\linewidth]{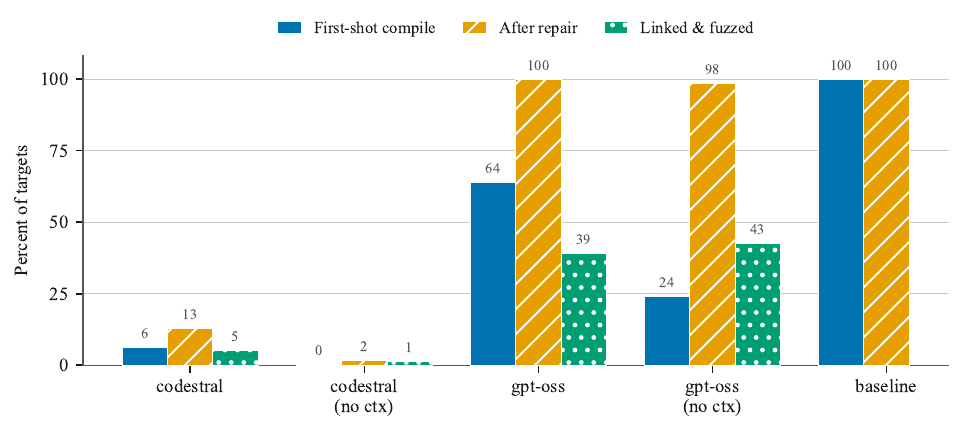}
  \caption{First-shot object compileability, post-repair object compileability, and the fraction of harnesses that linked and reached the fuzzer.}
  \label{fig:compile}
\end{figure}

Model choice strongly affected first-shot compileability. As shown in Table~\ref{tab:comp}, the reasoning model compiled \NCompiledGpt{} of \NTargets{} harnesses with static context, whereas the code-specialized model compiled only 46 and 3 of 740 harnesses (6\% and under 1\%) across its two conditions. Removing the source window and finding description reduced the reasoning model's rate from 63.9\% to 24.2\%, confirming the value of static context. Figure~\ref{fig:compile} shows that repair raised its object compileability to 100\% in both conditions after a mean of 1.2 rounds with context and 1.3 without; the code-specialized model improved only marginally.

\begin{figure}[!t]
  \centering
  \includegraphics[width=\linewidth]{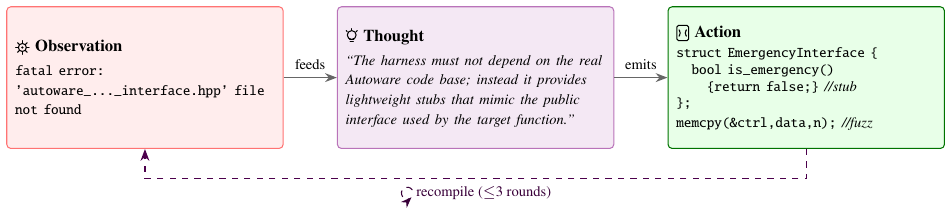}
  \caption{A compiler-guided repair iteration from a GPT-OSS harness. In response to a missing dependency, the model removes the real dependency and introduces interface stubs. Repeated repair makes the harness compile while moving it away from the intended Autoware implementation}
  \label{fig:react}
\end{figure}

This gain largely reflected stub convergence. As illustrated in Figure~\ref{fig:react}, the reasoning model often replaced unresolved dependencies with local stubs. Consequently, only \LinkedGpt{} of \NTargets{} context-enabled harnesses and \LinkedGptNo{} of \NTargets{} no-context harnesses linked and reached the fuzzer; 451 and 413 nominally compilable harnesses, respectively, failed to bind to Autoware symbols. The no-context condition linked more harnesses, 315 versus 289, because it produced simpler self-contained stubs. Repair therefore shifted the bottleneck from compilation to linking, making object compileability an unreliable proxy for valid analysis. The baseline reinforces this distinction: although every harness compiled, none exercised target logic, yielding average libFuzzer feature coverage of 2 compared with approximately 190 for the linked reasoning-model harnesses.

\subsection{Confirmation Analysis}
\begin{figure}[htbp]
  \centering
  \includegraphics[width=\linewidth]{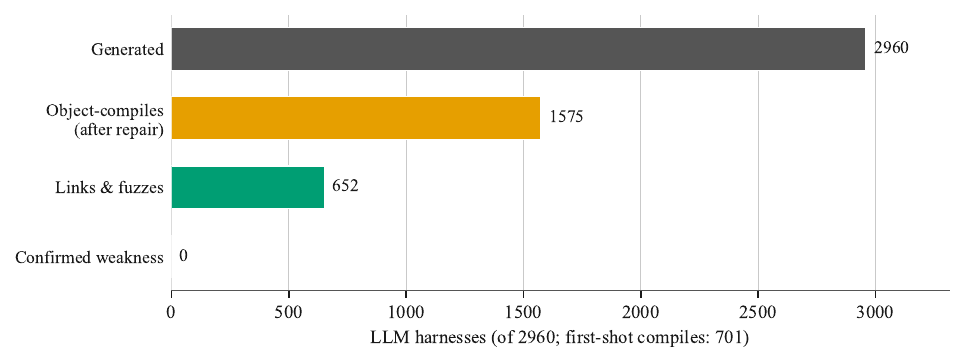}
  \caption{Attrition of LLM-generated harnesses across the four conditions}
  \label{fig:funnel}
\end{figure}

The outcomes depend on whether a harness reached and exercised the real target.
Figure~\ref{fig:funnel} summarizes this attrition. After repair, \NDisconfirmed{} LLM harnesses linked and fuzzed for the full budget without a crash: 24 and 10 from the code-specialized model (with and without static context), and 271 and 310 from the reasoning model (with and without context). These runs provide only weak disconfirmation because most surviving harnesses exercised local stubs rather than the intended Autoware implementation. The remaining 2{,}308 condition--target pairs never reached the fuzzer.

Four cases illustrate the aggregate results as shown in Table~\ref{tab:cases}. First, a P1 control-gate harness compiled and fuzzed cleanly only after replacing real interfaces with no-op stubs and reimplementing the input structure, so it did not exercise the reported Autoware branch. Second, the code-specialized model rarely produced compilable harnesses, mainly because of invalid target paths, missing headers, and leaked Markdown syntax. Third, static context increased the reasoning model's first-shot compileability from 24.2\% to 63.9\%, indicating that it improves artifact construction. Finally, a velocity-smoother crash occurred inside a model-reimplemented helper with no Autoware frame on the stack. Together, these cases show that compilation or crashing provides evidence only when the harness demonstrably executes the real target code.

Overall, the results show that the primary obstacle to LLM-assisted dynamic confirmation is not fuzz-input generation but faithful integration with the target software. Static context improves first-shot artifact quality, and compiler-guided repair improves nominal compileability, but neither reliably preserves linkage to the real implementation. For large, dependency-rich automated-driving stacks like Autoware, successful object compilation must therefore be separated from successful linking, target reachability, and genuine dynamic confirmation. The ablation in which static context raised the reasoning model's first-shot compileability from 24.2\% to 63.9\%, showing that the static stage improves artifact quality rather than merely selecting targets.

\begin{table}[!ht]
  \caption{Summary of the Four Case Studies}\label{tab:cases}
  \begin{center}\footnotesize
  \begin{tabularx}{\linewidth}{p{0.02\linewidth}p{0.24\linewidth}X}
    \toprule
    \# & Setting & Outcome \\
    \midrule
    1 & Control command gate, emergency arbitration (P1, control); gpt-oss with static context &
    Compiled, linked, and fuzzed for the full budget with no crash, but only by stubbing the real
    interfaces. A weak disconfirmation; compileability overstates how much real logic is exercised. \\
    2 & Code-specialized model, all 30 targets &
    Compiled 0/30, dominated by including the target \texttt{.cpp} and omitting ROS headers. Code
    specialization did not overcome cross-package build integration. \\
    3 & Static-context ablation (gpt-oss) &
    Static context raised first-shot compileability from 13.3\% to 56.7\%. The static stage
    materially improves artifact quality, not just target selection. \\
    4 & Lone crash, velocity smoother (gpt-oss) &
    The crash was an unchecked \texttt{std::vector::at} in the harness's own stub, not in Autoware.
    A crash is meaningless unless the harness provably drives the real code. \\
    \bottomrule
  \end{tabularx}
  \end{center}
\end{table}

\section{Discussion}\label{sec:discussion}

The observed failures are systemic rather than target-specific. They arise at the boundary between a syntactically plausible harness and the build graph of a large ROS~2 workspace, where successful integration depends on resolving package dependencies, generated message types, include paths, link targets, and node-construction requirements. This difficulty reflects the scale of the subject system, including hundreds of packages, \NIfaces{} interfaces, and a distributed validation surface, rather than the characteristics of any single candidate.

The results consistently support this conclusion. The strongest first-shot condition compiled only 473 of 740 harnesses, and 1,817 of the 2,259 initial failures were caused by dependency-wiring errors. Compiler-guided repair raised object compileability to 100\% for the reasoning model, yet only 652 of 2,960 LLM-generated harnesses ultimately linked and reached the fuzzer. Thus, compilation is not the principal assurance objective.

The repair loop fails because it optimizes for satisfying the compiler rather than preserving the connection to the real target. When faced with unresolved dependencies, the model often removes them and introduces local stubs, producing self-contained artifacts that compile but no longer exercise Autoware. The bottleneck therefore shifts from compilation to linking and target reachability rather than being removed. A repair process suitable for software assurance would need to optimize explicitly for integration with the native build. This would require resolving the correct package dependencies and build targets, linking against real implementation objects, and verifying at runtime that the intended Autoware functions appear on the executed path. These requirements extend beyond prompt refinement and call for build-aware tooling that treats semantic target preservation as a first-class constraint.

The results also limit the immediate role of LLM-generated harnesses in autonomous driving continuous integration. On current evidence, they are not yet reliable as an autonomous dynamic-assurance stage. They may nevertheless be useful for producing initial harness drafts that a human engineer completes and validates.



\section{Conclusions}\label{sec:conclusion}
This study evaluated whether LLMs can automate the dynamic confirmation of software weaknesses in an AV software stack. The results support a qualified negative. The static analysis identifies a broad safety-relevant attack surface in Autoware, comprising \NDecision{} decision rules, \NValidation{} validation checks, and \NFlows{} input-to-safety-output paths across \NPkgs{} packages. However, the dynamic stage shows that the main obstacle is not fuzzing, but faithful integration with the real build. Unaided model outputs rarely compile and link against a stack of this scale. Compiler-in-the-loop repair raises object compileability to 100\% for the stronger model, but largely through stub convergence. As a result, only 652 of 2,960 harnesses reach the fuzzer, and all 37 crashes occur in generated stub code rather than in Autoware. 

The results are qualified by several limitations. The decision-rule classification and package-level flow analysis are heuristic. As a result, their counts characterize the observed attack surface rather than provide a sound over- or under-approximation. Only packages with available compile commands contribute to the compiler-precise inventory. All experiments are conducted in software-in-the-loop; therefore, we make no claim about exploitability on a physical vehicle. The realized fuzzing time was \FuzzRealized{} of the configured \FuzzBudget{} seconds per target. Because few harnesses linked and reached execution, this difference does not affect the main finding. Compileability is defined as successful object compilation rather than full linking. Many compiling harnesses also replace real interfaces with local stubs. The reported compileability rates and disconfirmations are therefore optimistic upper bounds on useful analysis. Under a full-link and real-target-execution criterion, the effective success rate approaches zero across all conditions. The repair loop optimizes object compilation, uses a fixed budget of \RepairRounds{} rounds, and relies on the same model for generation and repair. A link-aware repair process should perform better, although the present design clearly exposes the stub-convergence failure mode. Autoware is public and may have appeared in model pretraining data, which would favor the models. Even so, the observed success rates remain low. Because the study evaluates only Autoware and two open-weight models, the reported rates should not be generalized directly. The broader hypothesis is that build integration, rather than input generation, is the binding constraint in large robotics and automated-driving codebases.

Future automation should therefore focus on dependency resolution, native linking, and verified execution of the intended target rather than on prompt refinement or compile-only repair. The static characterization remains valuable independently by prioritizing high-impact sites for human review. We release the prompts, generated artifacts, static-analysis outputs, and execution logs to support reproduction and further study.

\section{Acknowledgments}
We used the generative AI tools `ChatGPT' and `Claude' to help rephrase parts of our own writing to improve clarity and for editorial purposes.

\section*{AUTHOR CONTRIBUTIONS}
\textbf{Md. Wasiul Haque, Sagar Dasgupta:} conceptualization, methodology, coding, data collection, data analysis, and writing – original draft; \textbf{Mizanur Rahman:} conceptualization, methodology, writing – original draft, review and editing, and funding acquisition; \textbf{Md Rayhanur Rahman:} conceptualization, methodology, writing – original draft, review and editing.

\section*{DECLARATION OF CONFLICTING INTERESTS}
The authors declared no potential conflicts of interest with respect to the research, authorship, and/or publication of this article.

\section*{FUNDING}
This research was supported by the National Center for Transportation Cybersecurity and Resiliency (TraCR) (a U.S. Department of Transportation National University Transportation Center) headquartered at Clemson University, Clemson, South Carolina, USA (Award \# 69A3552344812, 69A3552348317) and National Science Foundation (NSF) (Award \# 2340456). Any opinions, findings, conclusions, and recommendations expressed in this material are those of the author(s) and do not necessarily reflect the views of funding agencies, and the U.S. Government assumes no liability for the contents or use thereof.


\newpage
\bibliographystyle{trb}
\bibliography{main}

@article{macenski2022ros2,
  author  = {Steven Macenski and Tully Foote and Brian Gerkey and Chris Lalancette and William Woodall},
  title   = {Robot Operating System 2: Design, architecture, and uses in the wild},
  journal = {Science Robotics},
  volume  = {7},
  number  = {66},
  pages   = {eabm6074},
  year    = {2022},
  doi     = {10.1126/scirobotics.abm6074}
}

@article{yurtsever2020survey,
  author  = {Ekim Yurtsever and Jacob Lambert and Alexander Carballo and Kazuya Takeda},
  title   = {A Survey of Autonomous Driving: Common Practices and Emerging Technologies},
  journal = {IEEE Access},
  volume  = {8},
  pages   = {58443--58469},
  year    = {2020},
  doi     = {10.1109/ACCESS.2020.2983149}
}

@misc{webb2020waymo,
  author       = {Nick Webb and Dan Smith and Christopher Ludwick and Trent Victor and Qi Hommes and Francesca Favar\`o and George Ivanov and Tom Daniel},
  title        = {Waymo's Safety Methodologies and Safety Readiness Determinations},
  year         = {2020},
  howpublished = {arXiv:2011.00054}
}

@misc{shalevshwartz2017rss,
  author       = {Shai Shalev-Shwartz and Shaked Shammah and Amnon Shashua},
  title        = {On a Formal Model of Safe and Scalable Self-driving Cars},
  year         = {2017},
  howpublished = {arXiv:1708.06374}
}

@inproceedings{gog2021pylot,
  author    = {Ionel Gog and Sukrit Kalra and Peter Schafhalter and Matthew A. Wright and Joseph E. Gonzalez and Ion Stoica},
  title     = {Pylot: A Modular Platform for Exploring Latency--Accuracy Tradeoffs in Autonomous Vehicles},
  booktitle = {Proceedings of the IEEE International Conference on Robotics and Automation (ICRA)},
  pages     = {8806--8813},
  year      = {2021}
}

@misc{apollo,
  author       = {{Baidu}},
  title        = {{Apollo}: An Open Autonomous Driving Platform},
  year         = {2024},
  howpublished = {\url{https://github.com/ApolloAuto/apollo}},
  note         = {Accessed 2026}
}

@misc{openpilot,
  author       = {{comma.ai}},
  title        = {openpilot: An Operating System for Robotics},
  year         = {2024},
  howpublished = {\url{https://github.com/commaai/openpilot}},
  note         = {Accessed 2026}
}

@inproceedings{kato2018autoware,
  author    = {Shinpei Kato and Shota Tokunaga and Yuya Maruyama and Seiya Maeda and Manato Hirabayashi and Yuki Kitsukawa and Abraham Monrroy and Tomohito Ando and Yusuke Fujii and Takuya Azumi},
  title     = {Autoware on board: Enabling autonomous vehicles with embedded systems},
  booktitle = {Proceedings of the 9th ACM/IEEE International Conference on Cyber-Physical Systems (ICCPS)},
  pages     = {287--296},
  year      = {2018}
}

@misc{autowareDocs,
  author       = {{Autoware Foundation}},
  title        = {Autoware Documentation},
  year         = {2026},
  howpublished = {\url{https://autowarefoundation.github.io/autoware-documentation/main/home/}},
  note         = {Accessed 2026}
}

@article{jung2025autowareapollo,
  author  = {Hee-Yang Jung and Dong-Hee Paek and Seung-Hyun Kong},
  title   = {Open-Source Autonomous Driving Software Platforms: Comparison of Autoware and Apollo},
  journal = {arXiv preprint arXiv:2501.18942},
  year    = {2025},
  doi     = {10.48550/arXiv.2501.18942}
}

@article{koopman2016challenges,
  author  = {Philip Koopman and Michael Wagner},
  title   = {Challenges in Autonomous Vehicle Testing and Validation},
  journal = {SAE International Journal of Transportation Safety},
  volume  = {4},
  number  = {1},
  pages   = {15--24},
  year    = {2016},
  doi     = {10.4271/2016-01-0128}
}

@techreport{kalra2016driving,
  author      = {Nidhi Kalra and Susan M. Paddock},
  title       = {Driving to Safety: How Many Miles of Driving Would It Take to Demonstrate Autonomous Vehicle Reliability?},
  institution = {RAND Corporation},
  year        = {2016},
  doi         = {10.7249/RR1478}
}

@article{dieber2017security,
  author  = {Bernhard Dieber and Benjamin Breiling and Sebastian Taurer and Severin Kacianka and Stefan Rass and Peter Schartner},
  title   = {Security for the Robot Operating System},
  journal = {Robotics and Autonomous Systems},
  volume  = {98},
  pages   = {192--203},
  year    = {2017},
  doi     = {10.1016/j.robot.2017.09.017}
}

@article{vilches2018rvss,
  author  = {V\'ictor Mayoral-Vilches and Endika Gil-Uriarte and Irati Zamalloa Ugarte and Gorka Olalde Mendia and Rodrigo Izquierdo Pis\'on and Laura Alzola Kirschgens and Asier Bilbao Calvo and Alejandro Hern\'andez Cordero and Lucas Apa and C\'esar Cerrudo},
  title   = {Towards an Open Standard for Assessing the Severity of Robot Security Vulnerabilities, the Robot Vulnerability Scoring System ({RVSS})},
  journal = {arXiv preprint arXiv:1807.10357},
  year    = {2018}
}

@misc{cvss31,
  author       = {{FIRST.org}},
  title        = {Common Vulnerability Scoring System version 3.1: Specification Document},
  year         = {2019},
  howpublished = {\url{https://www.first.org/cvss/v3.1/specification-document}}
}

@inproceedings{serebryany2012asan,
  author    = {Konstantin Serebryany and Derek Bruening and Alexander Potapenko and Dmitriy Vyukov},
  title     = {{AddressSanitizer}: A Fast Address Sanity Checker},
  booktitle = {Proceedings of the USENIX Annual Technical Conference (ATC)},
  pages     = {309--318},
  year      = {2012}
}

@inproceedings{fioraldi2020aflpp,
  author    = {Andrea Fioraldi and Dominik Maier and Heiko Ei{\ss}feldt and Marc Heuse},
  title     = {{AFL++}: Combining Incremental Steps of Fuzzing Research},
  booktitle = {Proceedings of the 14th USENIX Workshop on Offensive Technologies (WOOT)},
  year      = {2020}
}

@misc{libfuzzer,
  author       = {{LLVM Project}},
  title        = {libFuzzer -- a library for coverage-guided fuzz testing},
  year         = {2024},
  howpublished = {\url{https://llvm.org/docs/LibFuzzer.html}},
  note         = {Accessed 2026}
}

@inproceedings{serebryany2017ossfuzz,
  author    = {Kostya Serebryany},
  title     = {{OSS-Fuzz} -- Google's Continuous Fuzzing Service for Open Source Software},
  booktitle = {USENIX Security Symposium (invited talk)},
  year      = {2017}
}

@inproceedings{ispoglou2020fuzzgen,
  author    = {Kyriakos Ispoglou and Daniel Austin and Vishwath Mohan and Mathias Payer},
  title     = {{FuzzGen}: Automatic Fuzzer Generation},
  booktitle = {Proceedings of the 29th USENIX Security Symposium},
  pages     = {2271--2287},
  year      = {2020}
}

@inproceedings{deng2023titanfuzz,
  author    = {Yinlin Deng and Chunqiu Steven Xia and Haoran Peng and Chenyuan Yang and Lingming Zhang},
  title     = {Large Language Models Are Zero-Shot Fuzzers: Fuzzing Deep-Learning Libraries via Large Language Models},
  booktitle = {Proceedings of the 32nd ACM SIGSOFT International Symposium on Software Testing and Analysis (ISSTA)},
  pages     = {423--435},
  year      = {2023},
  doi       = {10.1145/3597926.3598067}
}

@inproceedings{xia2024fuzz4all,
  author    = {Chunqiu Steven Xia and Matteo Paltenghi and Jia Le Tian and Michael Pradel and Lingming Zhang},
  title     = {{Fuzz4All}: Universal Fuzzing with Large Language Models},
  booktitle = {Proceedings of the 46th IEEE/ACM International Conference on Software Engineering (ICSE)},
  year      = {2024},
  doi       = {10.1145/3597503.3639121}
}

@article{zhang2024fuzzdriver,
  author  = {Cen Zhang and Yaowen Zheng and Mingqiang Bai and Yeting Li and Wei Ma and Xiaofei Xie and Yuekang Li and Limin Sun and Yang Liu},
  title   = {How Effective Are They? Exploring Large Language Model Based Fuzz Driver Generation},
  journal = {Proceedings of the ACM SIGSOFT International Symposium on Software Testing and Analysis (ISSTA)},
  year    = {2024},
  doi     = {10.1145/3650212.3680355}
}

@misc{ossfuzzgen,
  author       = {{Google}},
  title        = {{OSS-Fuzz-Gen}: LLM-aided Fuzz Target Generation},
  year         = {2024},
  howpublished = {\url{https://github.com/google/oss-fuzz-gen}},
  note         = {Accessed 2026}
}

@inproceedings{meng2024chatafl,
  author    = {Ruijie Meng and Martin Mirchev and Marcel B\"ohme and Abhik Roychoudhury},
  title     = {Large Language Model Guided Protocol Fuzzing},
  booktitle = {Proceedings of the Network and Distributed System Security Symposium (NDSS)},
  year      = {2024},
  doi       = {10.14722/ndss.2024.24556}
}

@article{fang2024llmexploit,
  author  = {Richard Fang and Rohan Bindu and Akul Gupta and Daniel Kang},
  title   = {LLM Agents can Autonomously Exploit One-day Vulnerabilities},
  journal = {arXiv preprint arXiv:2404.08144},
  year    = {2024}
}

@inproceedings{cao2019adversarial,
  author    = {Yulong Cao and Chaowei Xiao and Benjamin Cyr and Yimeng Zhou and Won Park and Sara Rampazzi and Qi Alfred Chen and Kevin Fu and Z. Morley Mao},
  title     = {Adversarial Sensor Attack on {LiDAR}-based Perception in Autonomous Driving},
  booktitle = {Proceedings of the ACM SIGSAC Conference on Computer and Communications Security (CCS)},
  pages     = {2267--2281},
  year      = {2019},
  doi       = {10.1145/3319535.3339815}
}

@inproceedings{sato2021dirty,
  author    = {Takami Sato and Junjie Shen and Ningfei Wang and Yunhan Jia and Xue Lin and Qi Alfred Chen},
  title     = {Dirty Road Can Attack: Security of Deep Learning based Automated Lane Centering under Physical-World Attack},
  booktitle = {Proceedings of the 30th USENIX Security Symposium},
  pages     = {3309--3326},
  year      = {2021}
}

@inproceedings{cousot1977abstract,
  author    = {Patrick Cousot and Radhia Cousot},
  title     = {Abstract interpretation: A unified lattice model for static analysis of programs by construction or approximation of fixpoints},
  booktitle = {Proceedings of the Fourth Annual ACM SIGPLAN-SIGACT Symposium on Principles of Programming Languages (POPL)},
  pages     = {238--252},
  year      = {1977}
}

@misc{codeqlOverview,
  author       = {{GitHub}},
  title        = {{CodeQL}},
  year         = {2026},
  howpublished = {\url{https://codeql.github.com/}},
  note         = {Accessed 2026}
}

@techreport{iso26262,
  author      = {{International Organization for Standardization}},
  title       = {{ISO} 26262:2018 Road Vehicles --- Functional Safety},
  institution = {ISO},
  year        = {2018},
  address     = {Geneva, Switzerland}
}

@techreport{isosae21434,
  author      = {{International Organization for Standardization} and {SAE International}},
  title       = {{ISO/SAE} 21434:2021 Road Vehicles --- Cybersecurity Engineering},
  institution = {ISO/SAE},
  year        = {2021},
  address     = {Geneva, Switzerland}
}

@techreport{saej3016,
  author      = {{SAE International}},
  title       = {{J3016}: Taxonomy and Definitions for Terms Related to Driving Automation Systems for On-Road Motor Vehicles},
  institution = {SAE International},
  number      = {J3016\_202104},
  year        = {2021}
}

@article{haque2025security,
  title={Security Vulnerabilities in Software Supply Chain for Autonomous Vehicles},
  author={Haque, Md Wasiul and Erfan, Md and Dasgupta, Sagar and Rahman, Md Rayhanur and Rahman, Mizanur},
  journal={arXiv preprint arXiv:2509.16899},
  year={2025}
}

@article{manes2019fuzzing,
  author  = {Valentin J. M. Man\`es and HyungSeok Han and Choongwoo Han and Sang Kil Cha and Manuel Egele and Edward J. Schwartz and Maverick Woo},
  title   = {The Art, Science, and Engineering of Fuzzing: A Survey},
  journal = {IEEE Transactions on Software Engineering},
  volume  = {47},
  number  = {11},
  pages   = {2312--2331},
  year    = {2021},
  doi     = {10.1109/TSE.2019.2946563}
}

@article{hou2024llmse,
  author  = {Xinyi Hou and Yanjie Zhao and Yue Liu and Zhou Yang and Kailong Wang and Li Li and Xiapu Luo and David Lo and John Grundy and Haoyu Wang},
  title   = {Large Language Models for Software Engineering: A Systematic Literature Review},
  journal = {ACM Transactions on Software Engineering and Methodology},
  volume  = {33},
  number  = {8},
  year    = {2024},
  doi     = {10.1145/3695988}
}

@inproceedings{yao2023react,
  author    = {Shunyu Yao and Jeffrey Zhao and Dian Yu and Nan Du and Izhak Shafran and Karthik Narasimhan and Yuan Cao},
  title     = {{ReAct}: Synergizing Reasoning and Acting in Language Models},
  booktitle = {Proceedings of the International Conference on Learning Representations (ICLR)},
  year      = {2023}
}

@inproceedings{lattner2004llvm,
  author    = {Chris Lattner and Vikram Adve},
  title     = {{LLVM}: A Compilation Framework for Lifelong Program Analysis \& Transformation},
  booktitle = {Proceedings of the International Symposium on Code Generation and Optimization (CGO)},
  pages     = {75--88},
  year      = {2004},
  doi       = {10.1109/CGO.2004.1281665}
}

@misc{clangtooling,
  author       = {{LLVM Project}},
  title        = {Clang {LibTooling} and the {AST} Matcher Reference},
  year         = {2024},
  howpublished = {\url{https://clang.llvm.org/docs/LibTooling.html}},
  note         = {Accessed 2026}
}

@inproceedings{aggarwal2006integrating,
  title={Integrating static and dynamic analysis for detecting vulnerabilities},
  author={Aggarwal, Ashish and Jalote, Pankaj},
  booktitle={30th Annual International Computer Software and Applications Conference (COMPSAC'06)},
  volume={1},
  pages={343--350},
  year={2006},
  organization={IEEE}
}
\end{document}